# Synthesis and Characterization of Mg doped ZnFe$_2$O$_4$

*Kemi Y. Adewale[1,*], Itegbeyogene P.Ezekiel[2]*

[1]School of Chemistry and Physics, Westville Campus, University of KwaZulu-Natal, Durban, South Africa
[2]School of Chemistry and Physics, Westville Campus, University of KwaZulu-Natal, Durban, South Africa

*Abstract*

*Single-phase Mg-doped ZnFe2O4 nanoparticles with x= 0, 0.3, 0.5, 0.7 have been prepared by the glycol-thermal method without any subsequent calcination. The crystallite size, microstructure and magnetic properties of the prepared nanoparticles were studied using X-ray diffraction (XRD), high resolution scanning electron microscope (TEM), Mössbauer spectroscopy and vibrating sample magnetometer at room temperature. The XRD results revealed the production of a sharp single cubic spinel structure in all the synthesized samples without any impurity peak with the average crystallite size of about 19–28 nm. It was noticed that the lattice parameter varies as the Mg2+ ion concentration increases. 57Mössbauer measurement showed that the nano ferrites exhibit ferrimagnetic and superparamagnetic states. Magnetization measurements confirmed the superparamagnetic behaviour of the samples. The highest coercivity and saturation magnetization were observed at x=0.3. The saturation magnetization (MS) decreases while coercivity (HC) varies with an increase in the concentration of Mg2+ ion.*

*\*Author for Correspondence* E-mail: kemiadewale2@gmail.com

## INTRODUCTION

Nanocrystalline spinel ferrites are an important class of materials with varieties of electronic, magnetic and catalytic properties. Depending upon their properties, ferrite nanoparticles have numerous applications such as biomedical, ferrofluid, magnetic media, microwave, magnetocaloric refrigeration and gas sensors [1-8]. The subject of interest to study include structural, magnetic and electrical properties. Ferrites are important electronic and magnetic ceramics. Hence, their interest continues to expand [9]. It is known that the cationic distribution at tetrahedral (A) and octahedral (B) sites significantly affect properties of ferrites. The cationic distribution depends on various parameters such as synthesis techniques, processing conditions, sintering temperature, sintering time and chemical composition [10, 11].

Mixed zinc ferrites with high initial permeability are technically important. ZnFe$_2$O$_4$ has a normal spinel structure [12] and occupies tetrahedral (A) sites of the spinel lattice [13]. Composite phase of a spinel and hematite-like structure was observed for Ca$_x$Zn$_{1-x}$Fe$_2$O$_4$ when $x$=0.5 and hematite structure when $x$=1 [14]. Some reports describe magnesium ferrite (MgFe$_2$O$_4$) as having an inverse spinel structure with the Fe$^{3+}$ ions distributed between the tetrahedral (A) and octahedral (B) sites with most of the Mg$^{2+}$ ions located on B sites [15]. The aim of this work is to present a systematic study of the evolution of the magnetic properties of the system and to investigate how the properties change across the series.

## EXPERIMENTAL DETAILS

Mg$_x$Zn$_{1-x}$Fe$_2$O$_4$ ($x$ = 0, 0.3, 0.5, 0.7, and 1) were produced by the low temperature glycol-thermal process in a stirred pressure reactor. We used high purity calcium chloride (MgCl$_2$.6H$_2$O), zinc chloride (ZnCl$_2$) and ferric chloride (FeCl$_3$.6H$_2$O) as starting materials. An appropriate amount of each material in stoichiometric ratio is dissolved in a beaker of about 50 ml of de-ionized water and magnetically stirred. To initiate precipitation, excess aqueous ammonia solution (NH$_3$) was slowly added to the chloride mixture after about 10 minutes of stirring until the pH was about

9.4. The solution was further stirred for 40 minutes until complete precipitation was obtained. The precipitate was mixed with 250 ml of ethylene glycol and transferred into the reaction chamber of a Watlow PARR 4843 pressure reactor. The reactor was operated at a soak temperature of 200 °C for 6 hours at a pressure and speed of 170 psi and 300 rpm respectively. The obtained products were washed and filtered several times with de-ionized water over a Whatman filter in a Büchner flask in other to remove the chlorides. The addition of standard solution of silver nitrate ($AgNO_3$) to the filtrate was used to confirm the absence of chlorides. The cleaned samples produced were then dried under an infrared 250 W lamp for 24 hours and were finally homogenized using an agate mortar and pestle. The characterization of the samples was done at room temperature using a Co-Kα radiation X-ray diffraction (XRD), high-resolution scanning electron microscopy (HRSEM), $^{57}$Fe Mössbauer spectroscopy and vibrating sample magnetometer (VSM).

**RESULTS AND DISCUSSIONS**

Figure 1 shows the XRD patterns of $Mg_xZn_{1-x}Fe_2O_4$. All the peaks correspond to the cubic spinel structure and are correctly indexed. No impurity peaks were observed which imply that the samples made were all single phase. The average crystallite size for each sample were estimated using the Scherrer formula [16]. The calculated crystallite size, lattice parameters and the X-ray densities are presented in Table 1. The crystallite size and the lattice parameters vary for all the compositions of $x$. The fact that the ionic radius of $Zn^{2+}$ is bigger than the ionic radius of $Mg^{2+}$ might be responsible for that. The lattice parameter $a$ was obtained using Bragg's diffraction equation, $a = \frac{\lambda}{2\sin\theta}\sqrt{h^2+k^2+l^2}$ [17]. The values of the X-ray density $\rho_{XRD}$ have been obtained from the molecular weight and the volume using the equation $\rho_{XRD} = \frac{8M}{N_A a^3}$ where $M$ is the molecular weight of the samples, $N_A$ is the Avogadro's constant and $a$ is the lattice constant [18]. The number 8 in the formula is indicative that there are eight molecules per unit cell in the cubic spinel ferrite structure. Figure 2 shows the high-resolution scanning electron microscopy (HRSEM) micrographs of the samples. All the samples show spherically dispersed nanoparticles in shape across the surfaces and display a uniformly distribution of particles with different sizes.

***Table 1:** Crystallite sizes (D) for prepared samples of $Mg_xZn_{1-x}Fe_2O_4$ (x = 0, 0.3, 0.5, 0.7, and 1)*

| Sample | $D$ (nm) ±0.43 | $a$ (Å) ±0.006 | $\rho_{XRD}$ (g/cm$^3$) ±0.47 |
|---|---|---|---|
| 0 | 18.94 | 8.386 | 5.43 |
| 0.3 | 27.38 | 8.384 | 5.16 |
| 0.5 | 26.80 | 8.392 | 4.96 |
| 0.7 | 28.30 | 8.351 | 4.84 |
| 1 | 19.96 | 8.415 | 4.46 |

Room temperature Mössbauer spectra for $Mg_xZn_{1-x}Fe_2O_4$ nanoferrites are shown in Figure3. The fits were generated based on the Lorentzian site analysis using recoil Mössbauer analysis software. All the spectra were calibrated with respect to alpha-iron spectrum at room temperature. At least two sextets and one doublet were fitted to the spectra which represent the fraction of the Fe ions at both the A and B sites. One corresponds to $Fe^{3+}$ ions in the A sites and the other B sites with two different average nearest neighbor environments. The spectra $x$ = 0, 0.3 and 0.5 were best fitted to two sextets and two doublets. The sextets represent the $Fe^{3+}$ ions in ferrimagnetic state while the doublet indicates the $Fe^{3+}$ in a superparamagnetic state [19-23]. The sextet with the higher hyperfine magnetic field was assigned to the octahedral site due to the dipolar fields resulting from the deviation from cubic symmetry and from the covalent nature of the tetrahedral bonds [19, 21]. The doublets were assigned to A or B sites by the values of their isomer shifts. The spectra for $x$ =0.7 and $x$ = 1 were best fitted with two sextets and one doublet. Each doublet represents the paramagnetic state for the Fe at tetrahedral and octahedral sites. The lower isomer shifts on the A site than that of octahedral site is due to the higher covalency of the Fe-O bond on the tetrahedral site [23]. The various parameters obtained from the fitted Mössbauer spectra analysis including isomer shift ($\delta$), quadrupole splitting ($\Delta_{EQ}$), line widths (*LW*), fraction populations (*f*) and magnetic hyperfine fields ($B_{hf}$) are presented in Table2. The range of the values of the isomer shifts is consistent with $Fe^{3+}$ ions [22].

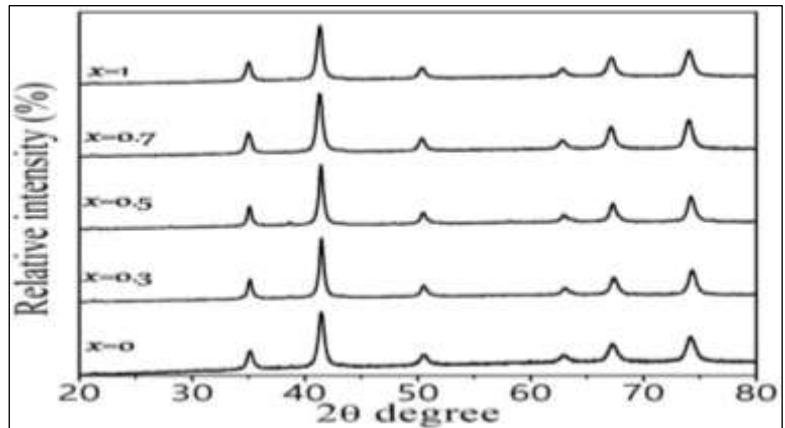

**Fig. 1:** *XRD patterns of* $Mg_xZn_{1-x}Fe_2O_4$ (*x* = 0, 0.3, 0.5, 0.7, and 1)

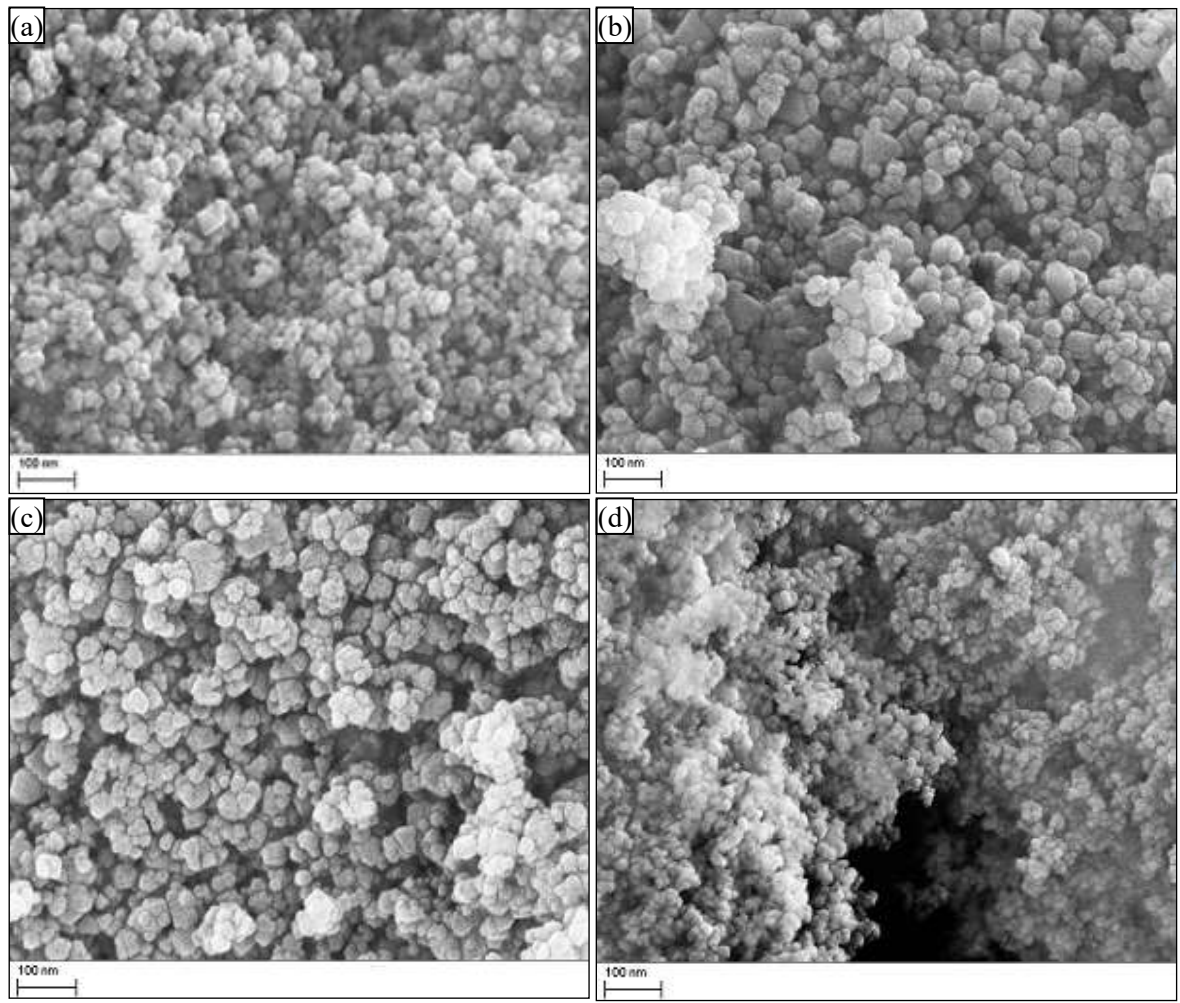

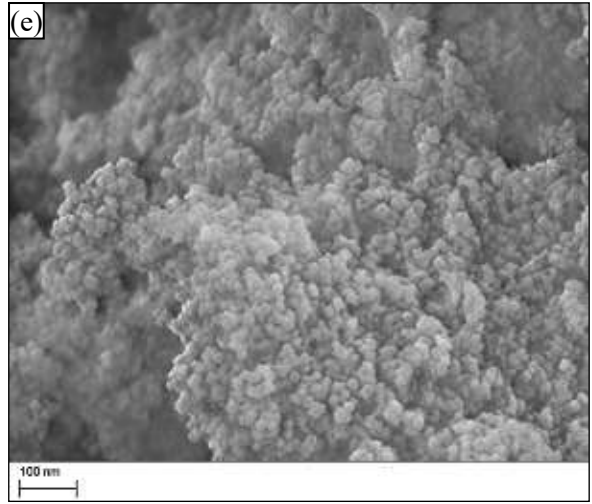

***Fig. 2:*** *HRSEM images of* $Mg_xZn_{1-x}Fe_2O_4$, (a) $x = 0$, (b) $x = 0.3$, (c) $x = 0.5$, (d) $x = 0.7$ and (e) $x = 1$.

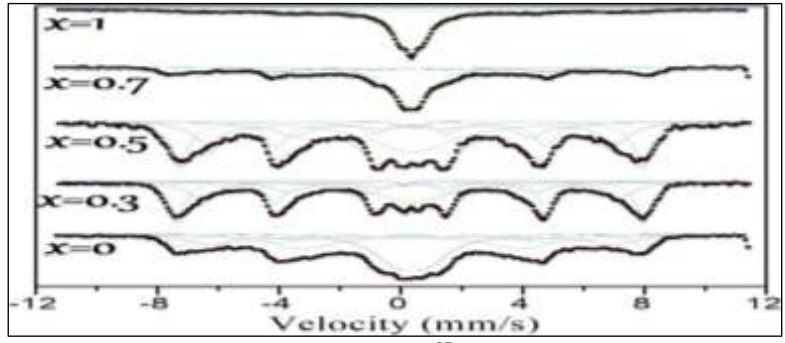

***Fig. 3:*** *Fitted room temperature $^{57}Fe$ Mössbauer spectra of* $Mg_xZn_{1-x}Fe_2O_4$ ($x = 0, 0.3, 0.5, 0.7,$ and $1$) *nanoferrites*.

***Table 2:*** *Isomer shift ($\delta$), quadrupole splitting ($\Delta_{EQ}$), line widths ($LW$), fraction populations ($f$) and magnetic hyperfine field ($B_{hf}$) of the samples* $Mg_xZn_{1-x}Fe_2O_4$ ($x = 0, 0.3, 0.5, 0.7,$ and $1$)

| $x$ | Pattern | $\delta$ (mm/s) | | $\Delta_{EQ}$ (mm/s) | | $LW$ (mm/s) | | $f$ (%) | | $B_{hf}$ (kOe) |
|---|---|---|---|---|---|---|---|---|---|---|
| | | ±0.003 | ±0.001 | ±0.01 | ±0.001 | ±0.08 | ±0.002 | ±0.03 | ±0.02 | ±0.11 |
| 0 | B | 0.37 | 0.31 | 7.22 | -0.002 | 1.40 | 0.37 | 30 | 15 | 470 |
| | A | 0.35 | 0.32 | 1.03 | 0.000 | 0.88 | 0.95 | 32 | 24 | 408 |
| 0.3 | B | 0.36 | 0.33 | 0.00 | 0.001 | 1.42 | 0.34 | 26.4 | 41.3 | 475.6 |
| | A | 0.12 | 0.33 | 0.00 | -0.019 | 0.11 | 0.57 | 0.59 | 31.7 | 428.2 |
| 0.5 | B | 0.38 | 0.33 | 0.00 | 0.004 | 1.72 | 0.42 | 31.2 | 36.2 | 471.4 |
| | A | 0.31 | 0.35 | 0.52 | 0.000 | 0.20 | 0.62 | 1.3 | 31.4 | 415.0 |
| 0.7 | B | 0.33 | 0.37 | 0.53 | -0.11 | 0.77 | 0.82 | 47 | 57 | 469 |
| | A | - | 0.42 | - | -0.05 | - | 0.70 | - | 17 | 277 |
| 1 | B | 0.35 | 0.46 | 0.43 | -0.05 | 0.51 | 1.17 | 63.1 | 24.2 | 423 |
| | A | - | 0.37 | 1.56 | -0.04 | - | 0.76 | - | 12.8 | 248 |

Figure 4 shows the room temperature hysteresis loops for the prepared $Mg_xZn_{1-x}Fe_2O_4$ nanoferrites in applied magnetic fields of up to 14 kOe. From the hysteresis loops, we calculated the magnetic parameters such as coercivity ($H_C$), saturation magnetization ($M_S$) remanent magnetizations ($M_R$) and squareness of the loops ($M_R/M_S$). The results are presented in Table 3. The coercive field is defined as

$$H_C = \left|\frac{H_{C1} + H_{C2}}{2}\right|$$

The $M_S$ were estimated from the law of approach to saturation magnetization using

$$M(H) = M_S\left(1 - \frac{a}{H} - \frac{b}{H^2}\right) + \chi H$$

where $M_S$, $a$, $b$ and $\chi$ are the best fit parameters to the data [24]. We have plotted the coercive fields, saturation magnetizations and remanent magnetizations against $x$ in Figures 5, 6 and 7.

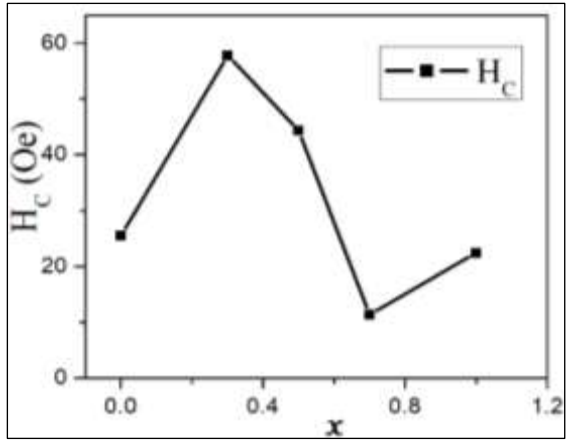

**Fig. 4:** *Variation of coercive field $H_C$ with $x$.*

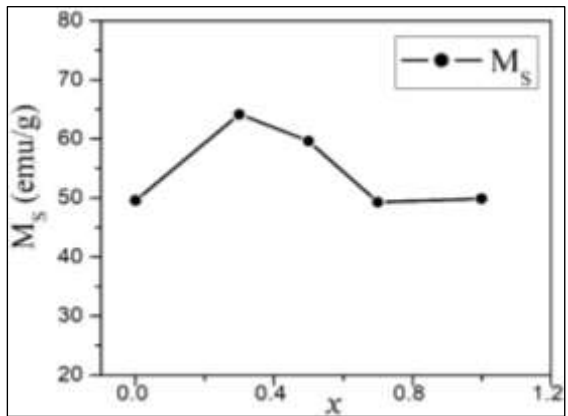

**Fig. 5:** *Variation of saturation magnetization $M_S$ with $x$.*

The variation of the calculated magnetic parameters is shown in Table 3. The coercive field $H_C$ ranges from 11.4 Oe to 57.8 Oe. $ZnFe_2O_4$ is a soft magnetic material, and when $Zn^{2+}$ in $ZnFe_2O_4$ is substituted by $Mg^{2+}$ ions, there is a lot of change in the magnetic properties like coercivity ($H_C$), saturation magnetization ($M_S$), and remanent magnetization ($M_R$) as shown in Table 3 [25, 26]. We obtained $K$ from Stoner-Wohlfaith model

$$H_C = \frac{0.96K}{M_S}.$$

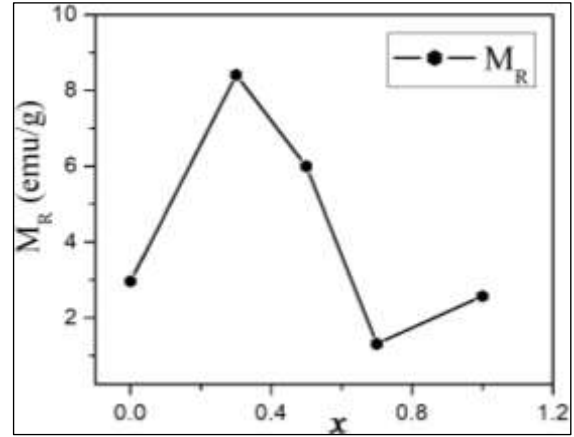

**Fig. 6:** *Variation of remanent magnetization $M_R$ with $x$.*

We observed that the $M_S$ varies for all the compositions of $x$. Hence, the $M_S$ of $Mg_xZn_{1-x}Fe_2O_4$ nanoparticles depends on the distribution of $Fe^{3+}$ ions among tetrahedral and octahedral lattice sites, because both $Mg^{2+}$ and $Zn^{2+}$ ions are non-magnetic in nature. The samples $Mg_{0.5}Zn_{0.5}Fe_2O_4$, $Mg_{0.7}Zn_{0.3}Fe_2O_4$ and $MgFe_2O_4$ with higher Mg concentration ($x = 0.5$, 0.7 and 1) showed more superparamagnetic behavior. This indicates that $Mg^{2+}$ ions occupied the tetrahedral sites and $Fe^{3+}$ ions at the octahedral sites, whereas the dopant, $Zn^{2+}$ ions occupied either octahedral or tetrahedral sites [27], thus showing superparamagnetic behavior. Similar results were reported in the literature [28]. The coercive field $H_C$ increases from 25.6 Oe for $ZnFe_2O_4$ to 57.8 Oe for $Mg_{0.3}Zn_{0.7}Fe_2O_4$. Then decreases to 44.3 Oe for $Mg_{0.5}Zn_{0.5}Fe_2O_4$ and to 11.4 Oe for $Mg_{0.7}Zn_{0.3}Fe_2O_4$. However, there is an increase from 11.4 Oe to 22.4 Oe in the field for $MgFe_2O_4$. This may be due to the increase of the super exchange interactions [29]. The variation of coercivity with the concentration of the metal ion can also be explained based on anisotropy field and domain wall energy [30]. The decrease of $H_c$ with decreasing grain size of the Zn composition may be attributed to the exchange coupling between the grains because it induces cooperative demagnetization processes starting from misaligned grains. For multi-domain systems the coercivity varies with the grain size [31]. The saturation magnetization $M_S$ followed almost the same trend with that of $H_C$. $Mg_{0.3}Zn_{0.7}Fe_2O_4$ has the highest value of saturation magnetization of 64 emu/g. On further increasing the concentration of Mg in $ZnFe_2O_4$ nanocrystals, the $M_S$ value of the samples $Mg_{0.5}Zn_{0.5}Fe_2O_4$ and

$Mg_{0.7}Zn_{0.3}Fe_2O_4$ decreased (i.e. 59.6 and 49.3 emu/g, respectively) (Figure 5). It is due to the $Mg^{2+}$ ions, which prefers to occupy the octahedral B sites which may push $Fe^{3+}$ to the tetrahedral sites, which in turn decreases the value of $M_S$. However, the migration towards A sites, would lead to the increase of $Fe^{3+}$ concentration in A sites, which gives rise to antiparallel spin coupling and spin canting, resulting in the weakening of AB exchange coupling, and thereby decreases the net magnetic moment [32-34]. The concentration and types of cations substitution also have very dominant effect on the magnetic properties. The net magnetic moment determines the $M_S$ value. Zinc ferrite ($ZnFe_2O_4$) is a normal spinel, non-magnetic ion $Zn^{2+}$ and magnetic ion $Fe^{3+}$ are distributed in A and B sites, respectively [35]. From the literature [36-38], the bulk magnesium ferrite ($MgFe_2O_4$) has an inverse spinel structure with the preference of $Mg^{2+}$ cations Mg-doped Zn ferrite, $Zn^{2+}$ and $Mg^{2+}$ ions prefer to inhabit the A and B sites, respectively, while $Fe^{3+}$ prefers to inhabit both A and B sites. When the content of Mg is larger than 0.5, $M_S$ decreased (Figure 6 and 7). This is due to the A-B exchange interaction, which is weaker than the B-B interaction. Moreover, the low values of $H_C$ and $M_R$ obtained for all the compositions indicate soft magnetic properties of the compounds [39-42].

**Table 3:** *Coercivity $H_C$, saturation magnetization $M_S$, remanence magnetization $M_R$ and squareness ratio ($M_R/M_S$) obtained at room temperatures in applied field of 14 kOe for $Mg_xZn_{1-x}Fe_2O_4$ (x = 0, 0.3, 0.5, 0.7, and 1) (Figure 7)*

| $x$ | $H_C$ (Oe) ±0.5 | $M_S$ (emu/g) ±0.3 | $M_R$ (emu/g) ±0.4 | $M_R/M_S$ ±0.05 | $K$ ±0.04 |
|---|---|---|---|---|---|
| 0 | 25.597 | 49.5 | 3.0 | 0.06 | 1.32 |
| 0.3 | 57.809 | 64.1 | 8.4 | 0.13 | 3.86 |
| 0.5 | 44.322 | 59.6 | 6.0 | 0.10 | 2.87 |
| 0.7 | 11.397 | 49.3 | 1.3 | 0.03 | 0.61 |
| 1 | 22.422 | 49.8 | 2.6 | 0.05 | 1.16 |

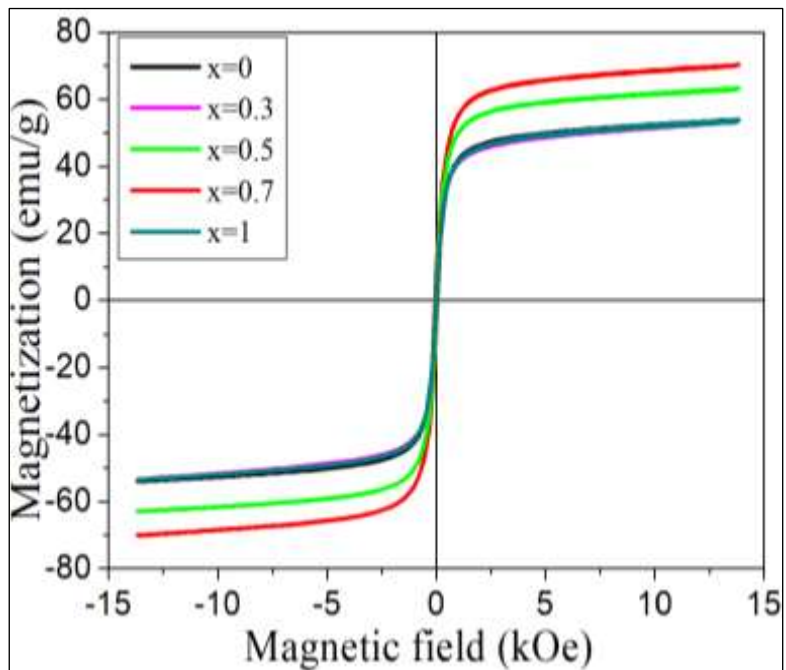

***Fig. 7:*** *Room temperature hysteresis loops of $Mg_xZn_{1-x}Fe_2O_4$ (x = 0, 0.3, 0.5, 0.7, and 1) nanoferrites.*

**CONCLUSIONS**
Mg-doped $ZnFe_2O_4$ ($Mg_xZn_{1-x}Fe_2O_4$) nanoferrites with x= 0, 0.3, 0.5, 0.7 and 1 were successfully synthesized by glycol-thermal method. XRD results confirmed single phase cubic spinel structure in all the synthesized samples without any impurity peak. The crystallite sizes were in the range of 18.9 - 28.3 nm. The obtained lattice parameter reduced from 8.384 ± 0.0059 to 8.351 ± 0.0061 Å by increasing the Mg content. HRSEM confirmed uniformly distributed particles with different sizes. Mössbauer spectroscopy results, recorded at room temperature showed that the samples exhibit ferrimagnetic and superparamagnetic states. Magnetization measurements at room temperature confirmed the superparamagnetic behavior of the samples. It is interesting to note that starting with $ZnFe_2O_4$, slightly doping with Mg causes a drastic change in the saturation magnetization. Magnetization results show a peak in the saturation magnetizations at $x = 0.3$.

**ACKNOWLEDGEMENTS**
The authors express gratitude to the National Research Foundation and the University of KwaZulu-Natal, South Africa for research grants.